\documentclass[12pt]{article}
\title{\bf On leptonic decay of a heavy quarkonium with a Higgs-boson emission}
\author{ G.A. Kozlov
\\
\em Bogoliubov Laboratory of Theoretical Physics,\\
\em Joint Institute for Nuclear Research,\\
\em 141980 Dubna, Moscow Region, Russia\\
\em e-mail: kozlov@jinr.ru\\
 }

\usepackage[dvips]{graphicx}

\begin{document}

\maketitle
\begin{abstract}

A leptonic $(\bar l l)$ decay of a heavy quark-antiquark bound state
 $T(\bar QQ)$ with a Higgs-boson $H$ emission is investigated. The
applying of the well-known low-energy theorem to meson-Higgs coupling
allows one to estimate the probability of the decay
$T(\bar QQ)\rightarrow \bar ll H$. The only a simple version of the
Standard Model extension containing two-Higgs doublet is considered.


\end{abstract}


\vspace {1 cm}



It is well-known that some extensions of the Standard Model (SM) admit the existence
of new physical bound states (hadrons), composed of heavy quarks and antiquarks including
the 4th generation quarks ($Q_4$) [1].\\
 The question of the existence of
4th generation fermions ($f_4$) is among the most important, intriguing and not solved yet one in the
modern elementary particle physics. We know that, e.g., the heterotic string phenomenology
in $E_{6}$ model leads to the 4th generation of leptons ($l_4$) and $Q_4$ with a
relatively stable massive neutrino of 4th generation ($\nu_4$) [2].
A possible virtual contributions of 4th generation particles have been advocated
by recent analysis [3] of precision data on the SM parameters.
The following question arises: what about the recent limits on the masses of $f_4$ ?
It turns out that $l_4$ and $Q_4$ are not excluded under
the condition that the Dirac $\nu_4$ is a (quasi)stable particle and it has a mass
around 50 GeV [3], and the rest of a spectrum of $f_4$-particles satisfies their direct experimental
constraints on the masses $m_4$ on the level above 80-220 GeV.
At the moment, the best
result on the lower bound restriction on $m_4$ was given by the CDF Collaboration at the Fermilab Tevatron,
using the measurement of the energy loss $dE/dx$ in a "calorimeter".
For the $up$-type $Q_4$ (labeled as $U$)
with the electric charge $e_{U}=+2/3$ this limit is $m_{U} >$ 220 GeV [4], that, in principal,
corresponds to the production cross-section of the order of $1~pb$ at the Tevatron energy.
It has been already reported [5], that in spite of the multi-$fb^{-1}$ luminosity which one expects
the Tevatron CDF and D0 to collect by the time the LHC will start, the
rates for heavy quarks will allow their abundant production already with typical start-up
luminosities of 1 $\%$ of the design, namely  $~ 10^{-32} cm^{-2} s^{-1}$.
The estimation leads to that the rate for pairs of heavy quarks
production at the LHC with the mass $~ O(400 GeV)$ is more than 100 times larger
than at the Tevatron.
In paper [6], we have already investigated
the issues of production and decays of hadrons containing the so-called light $Q_4$ with
the masses exceeded the top-quark mass, $m_{4} > m_{t}$.
We considered strongly bound states, made out of heavy quarks (including fourth family)
and using Higgs fields to bind them. There is an important special feature, because
unlike the exchange of gauge fields, the scalar particles attract both particles
and antiparticles, and the attraction of quarks by Higgs exchange is
independent of color. The scenario on the hypothesis that a bound state
can be formed from 6 top quarks and 6 anti-top quarks, held together
mainly by Higgs particle exchange, has been considered in [7].

Since there is no direct indications on the existence of the stable
$f_4$-fermions  (that means their small lifetime compared with the lifetime of the Universe)
it means, obviously, that one of the ways to explore these new particles is their search
for via the production and their identification through the decays at modern hadron colliders.
We assume that hadrons composed of $Q_4$-quarks are unstable and  effectively decay where
one of the final states should be the Higgs-boson. The reason of the Higgs-emission
is covered by the more probable and effective couplings between the Higgs-boson with heavy quarks.

In this letter, we consider the process of the Higgs-boson emission in decays
 $T(\bar QQ)\rightarrow V^{\star} H\rightarrow \bar ll H$,
where $T(\bar QQ)$ is the spin-1 heavy particle and $V^{\star}$ is a set of intermediate
neutral vector bosons including new generations of gauge bosons  (e.g., from $E_{6}$-model,
Little Higgs model [8]). Assuming an infinitely small momentum of the Higgs-boson when the
Higgs field is considered as the external one and does not carry the dependence on
the coordinates (the low-energy theorem [9]), the probability of the decay
 $T(\bar QQ)\rightarrow \bar ll H$ normalized to the Drell-Yan process
$T(\bar QQ)\rightarrow \bar ll$ is given by the formula:
\begin{eqnarray}
\label{e1}
R_{T(\bar QQ)\rightarrow \bar ll H/\bar ll}\equiv
\frac{\Gamma (T(\bar QQ)\rightarrow \bar ll H)}{\Gamma (T(\bar QQ)\rightarrow \bar ll )}=
\int^{s^{max}_{l}}_{0} ds_{l} \frac{\lambda^{1/2}(m^2_{T},m^2_{H},s_{l})}{24\pi^{2}v^2s_{l}}\eta^2_{HQ}\cr
{\left (1-\frac{4m^2_{l}}{s_{l}}\right )}^{1/2}\left (1+\frac{2m^2_{l}}{s_{l}}\right )
\frac{\lambda (m^2_{T},m^2_{H},s_{l}) + 6 m^2_{T} s_{l}}{{(m^2_{T}-s_{l})}^{2} +
\Gamma^{2}_{T} m^{2}_{T}},
\end{eqnarray}
where $s_{l} = (p_{l}+p_{\bar l})^{2}= 2(m^{2}_{l} + p_{l}\cdot p_{\bar l})$ is the
invariant mass of $\bar l l$-pair; the constants $\eta_{HQ}$ are defined for up (U)- and
down (D)- types of quarks in the form [1]
\begin{eqnarray}
\label{e2}
\eta_{HU}=\frac{\cos\alpha}{\sin\alpha} = \sin(\beta -\alpha) +
\cot\beta \cos(\beta-\alpha)\, ,
\end{eqnarray}

\begin{eqnarray}
\label{e3}
\eta_{HD}= - \frac{\sin\alpha}{\cos\beta} = \sin(\beta -\alpha) -
\tan\beta \cos(\beta-\alpha)\, .
\end{eqnarray}

In the decoupling regime reflecting the special ratio between the masses
of $Z$-boson ($m_{Z}$)  and CP-odd Higgs-boson $A$ ($m_{A}$),
$z=(m_{Z}/m_{A})^{2}<< 1$, the relations (\ref{e2}) and
(\ref{e3}) transform in the following distributions on the angle $\tan\beta = v_{U}/v_{D}$
for two vacuum expectation values $v_{U}$ and $v_{D}$:
\begin{eqnarray}
\label{e4}
\eta_{HU}\simeq 1+ z\, \sin(2\beta)\cos(2\beta)\tan^{-1}(\beta)\, ,
\end{eqnarray}

\begin{eqnarray}
\label{e5}
\eta_{HD}= 1- z\, \sin(2\beta)\cos(2\beta)\tan(\beta)\, .
\end{eqnarray}
The production rate for a light CP-even Higgs-boson can be estimated.
For illustration we plotted in Fig.1 in detail the $\sqrt{s_{l}}$-dependence
of $F_{T(\bar QQ)\rightarrow \bar ll H/\bar ll} = \Gamma^{-1} (T(\bar QQ)\rightarrow \bar ll )
d\Gamma (T(\bar QQ)\rightarrow \bar ll H)/d s_{l}$ on different values of $T(Q_{4}\bar Q_{4}$-
heavy quarkonia masses for $\tan\beta = 5$ and $\tan\beta = 20$ at fixed values
of the lightest CP-even Higgs-boson mass
$m_{H}= m_{h} = $120 GeV and $m_{A} = $300 GeV,
obeying the decoupling regime above mentioned.

We found that for $T(\bar U U)$-bound state the changing of
$F_{T(\bar QQ)\rightarrow \bar ll H/\bar ll}$ is very small with increasing of $m_{A}$ from
200 GeV up to 300 GeV at $5\leq\tan\beta\leq 20$. However, the situation changes drastically
if one considers the bound state composed of $D$-quarks. The amplitude
$F_{T(\bar QQ)\rightarrow \bar ll H/\bar ll}$ falls down with increasing of $M_{A}$.

In Fig. 2 we plotted the relative decay width
$R_{T(\bar QQ)\rightarrow \bar ll H/\bar ll}$ (\ref{e1}) versus the $T$-bound state mass $m_{T}\simeq 2 m_{Q}$
for $T(\bar U U)$- and $T(\bar D D)$- bound states, respectively at different $\tan\beta$.


\begin{figure} [htpb]
\begin{center}

\includegraphics[width=12cm,totalheight=8cm]{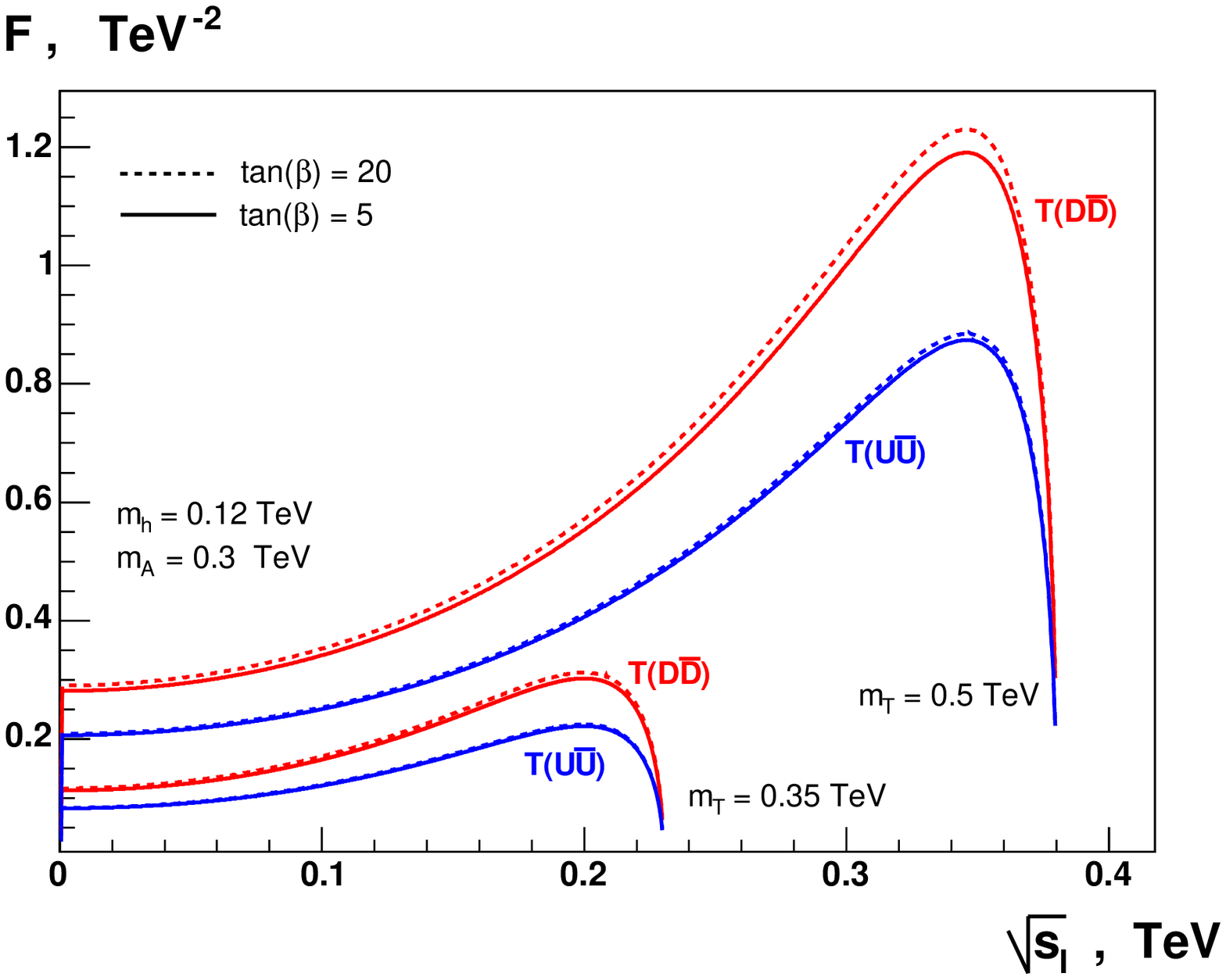}

Fig.1 {The differential distribution $F= \Gamma^{-1} (T(\bar QQ)\rightarrow \bar ll )
d \Gamma (T(\bar QQ)\rightarrow \bar ll H)/{d s_{l}}$ over invariant mass of the
lepton pair $s_{l}$ for
$T(U\bar U)$- and  $T(D\bar D)$- bound states
with different $\tan\beta = 5$ and $\tan\beta = 20$ at fixed $m_{h}=120$ GeV
and $m_{A}=300$ GeV as a function of $\sqrt{s_{l}}$.}
\vspace {1 cm}

\includegraphics [width=12cm,totalheight=9cm]{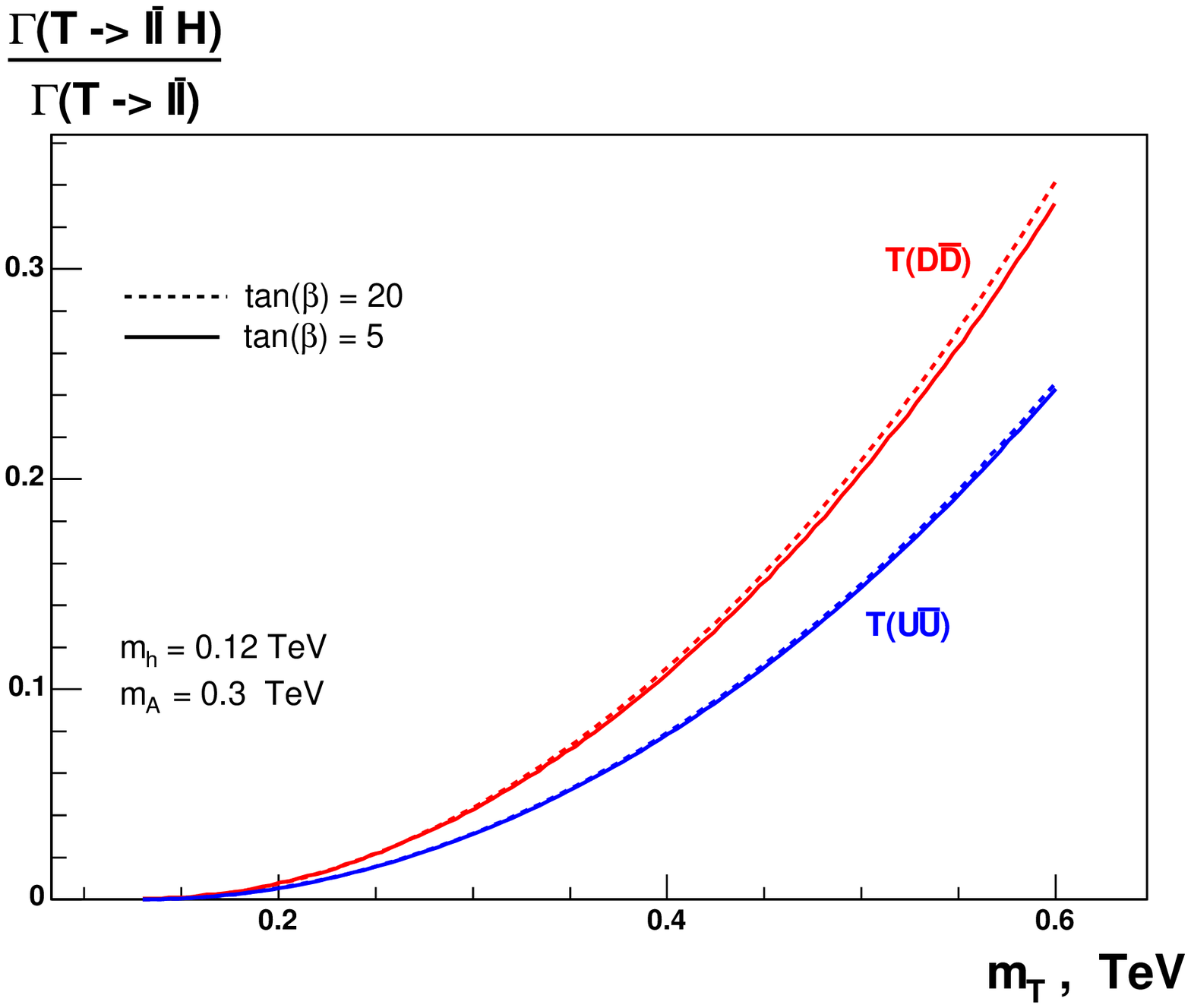}

Fig.2 The relative decay width $R$ (\ref{e1}) for $T(U\bar U)$- and
$T(D\bar D)$- bound states
for different $\tan\beta = 5$ and $\tan\beta = 20$ at fixed $m_{h}=120$ GeV
and $m_{A}=300$ GeV as a function of $m_{T}\simeq 2 m_{Q}$.

\end{center}
\end{figure}


\newpage


No essential difference are found with increasing of $\tan\beta$ from 5 to 20.
 For conclusion, the decays of heavy quarkonia are very good place to search for
a Higgs-boson in the light sector (e.g., CP-even $h$-boson). The decays we
have discussed, $T(\bar QQ)\rightarrow \bar ll H$, have branching ratios which
could potentially be probed by precision measurements at hadron colliders.
On the other hand, since there are three-body decays, the measurements of the
invariant mass spectra of leptons recoiling against a Higgs-boson may give
valuable insight into the dynamics of heavy quark-antiquark bound state involved.

\end{document}